\documentclass[12pt]{iopart}

\usepackage{epsfig,iopams,varioref}

\begin{document}


\title{Two-component boson systems with hyperspherical coordinates}

\author{T Sogo, O S{\o}rensen, A S Jensen and D V Fedorov}

\address{Department of Physics and Astronomy,
         University of Aarhus, DK-8000 Aarhus C, Denmark }

\ead{taka@phys.au.dk}

\begin{abstract}
The effective potential is computed for two boson systems in one trap
as a function of their two individual hyperadii and the distance between
their centers. Zero-range interactions are used and only relative
$s$-states are included. Existence and properties of minima are
investigated as a function of these three collective coordinates.
For sufficiently strong repulsion stable structures 
are found at a finite distance between the centers. 
The relative center of masses motion
corresponds to the lowest normal mode. The highest normal mode is
essentially the breathing mode where the subsystems vibrate by scaling
their radii in phase. The intermediate normal mode corresponds to
isovector motion where the subsystems vibrate by scaling their radii
in opposite phase. Stability conditions are established as
substantially more restrictive than in mean-field computations.
\end{abstract}

\submitto{\jpb}


\maketitle

\section{Introduction}

Condensed states of identical bosons at zero temperature in a
confining external field are accessible for experiments in many
laboratories \cite{pet01,pit03}.  Most often the condensates are
formed by one species of bosons, but also two-component structures have
been investigated.  The two components may consist of the same atom in
two different spin states as studied for $^{87}$Rb \cite{MBG97,HME98}.
The spin states may then be converted from one to the other while
maintaining the total number of atoms.  Also collective oscillations
can be studied experimentally as in \cite{MMF00} for $^{87}$Rb.
Two-component boson condensates of different atoms have other
properties as investigated by combining $^{87}$Rb and $^{41}$K
\cite{MMR02}.  More recently mixtures of boson and fermion
systems have been realized \cite{tru01,sch01,roa02,mod02,had02}.

The theory for two-component condensates are mostly based on the
mean-field approximation \cite{pet01,pit03}. The additional degrees of
freedom lead to spatial symmetry breaking in a cylindrical trap as
shown in \cite{CA99,EG99}.  Collective excitations in such traps are
predicted \cite{GS98}.  With two different trap frequencies also
spherical traps allow collective excitations \cite{PB98}.  The spatial
asymmetry is sometimes favored even in spherical traps \cite{Oh98}.
Repulsion between the two components can lead to separation of the two
components both in the ground state and for the collective oscillations
\cite{SC03}.  Stability conditions can be derived from properties of
the collective modes \cite{BCP97}.

Another type of formulation using hyperspherical coordinates for
one-component boson systems was introduced in \cite{boh98}.  This
formulation was extended to include two-body
correlations for one-component $N$-body boson systems \cite{sor02}.
This treatment allowed both very large scattering lengths and
attractive two-body interactions. The mathematical collapse is
prevented by a finite range interaction.  Conditions for the physical
collapse are rather similar to mean-field results. Still the computed
correlated structures for large scattering length may be physically
interesting \cite{sor02a}.

The present paper follows up with more details and new calculations on
a recent attempt to extend the hyperspherical formulation to
two-component boson systems \cite{sog04}. Although several new
features appear, we still only employ the lowest-order approximation
with inclusion of three collective coordinates. This is the minimum
number of degrees of freedom which can describe individual sizes of
the two subsystems and their separation distance.  A demand for more
complicated extensions may arise, but the specific direction is better
decided after digestion of the results from the simplest model survey.

In section \ref{sec2} we formulate the theoretical framework where the
effective potential is introduced. The properties of this key quantity
is investigated in details in section \ref{sec3}, especially with respect to
minima and their curvatures. Then we are equipped to establish
stability conditions for the total system as discussed in section \ref{sec4}.
Our conclusions are contained in section \ref{sec5}. 
Finally, for
completeness we include a number of rather tedious mathematical details in three
appendices, but the main text can be read independently because only
analytic and numerical results are used.

\section{Theoretical formulation}
\label{sec2}

We consider a dilute system of two species of bosons, $A$ and $B$,
with boson masses $m_A$ and $m_B$ and particle numbers $N_A$ and $N_B$.  We
label the particles with $i=1,2,3,...,N_A$ for compontent A 
and $i=N_A+1,N_A+2,...,N\equiv N_A+N_B$ for component B.  
The individual masses and position vectors are denoted 
$m_i$ and $\vec r_{i}$ where $i=1,2,3,...,N$.  These bosons are
trapped in a spherically symmetric harmonic oscillator potential with
the same trap frequency $\omega$.  The Hamiltonian describing this
system is
\begin{eqnarray}
\label{eq:TH}
\fl
H= H_0 + V_{A}+V_{B}+V_{AB} \\
\fl
H_0=\sum_{i=1}^{N_A}
[-\frac{\hbar^2}{2m_A}\vec \nabla^2_{i}+\frac{1}{2}m_A\omega^2 r_{i}^2] 
+\sum_{i=N_A+1}^{N}
[-\frac{\hbar^2}{2m_B}\vec \nabla^2_{i}+\frac{1}{2}m_B\omega^2 r_{i}^2] \\
\fl
V_{A}=\sum_{i>j=1}^{N_A}g_{A}\delta^{(3)}(\vec r_{ij}) \; ,\;\;
g_{A}=\frac{4\pi\hbar^2a_{A}}{m_A} \;, \\
\fl
V_{B}=\sum_{i>j=N_A+1}^{N}g_{B}\delta^{(3)}(\vec r_{ij}) \;, \;\;
g_{B}=\frac{4\pi\hbar^2a_{B}}{m_B} \;, \\ \label{e7}
\fl
V_{AB}=\sum_{i=1}^{N_A}\sum_{j=N_A+1}^{N}g_{AB}\delta^{(3)} (\vec r_{ij}) 
,\; g_{AB}=\frac{2\pi\hbar^2a_{AB}}{\mu_{AB}}  ,
\end{eqnarray}
where $\vec r_{ij} \equiv \vec r_{i} - \vec r_{j}$ denotes the vector
distance between two particles and the reduced mass of an $A$ and a
$B$ particle is given by $\mu_{AB}\equiv m_A m_B/(m_A+m_B)$.  The
interaction strengths of the two-body zero-range pseudo potentials,
$g_{A}$, $g_{B}$, and $g_{AB}$, are related to the $s$-wave scattering
lengths, $a_{A}$, $a_{B}$, and $a_{AB}$. This normalization is adopted
to provide the correct large-distance behavior of the effective
potential in mean-field computations \cite{sor04}.

We shall use hyperspherical coordinates with hyperradii defined for
both the individual and the total systems \cite{boh98,sor02}, i.e.
\begin{eqnarray} \label{e10}
 m \rho^2 \equiv \frac{1}{M} \sum_{i<j}^N m_i m_j r_{ij}^2 =
 \sum_{i=1}^N m_i r_i^2 - M R^2 \;, & \\ \label{e15} \rho_A^2 \equiv
 \frac{1}{N_A} \sum_{i<j}^{N_A} r_{ij}^2 \; , \;\; \rho_B^2 \equiv
 \frac{1}{N_B} \sum_{N_A<i<j}^{N} r_{ij}^2 \; ,& \\
\label{e20}
 \vec R_A = \frac{1}{N_A}\sum_{i=1}^{N_A}  \vec r_i   \; , \;\;
 \vec R_B = \frac{1}{N_B} \sum_{i=N_A+1}^{N} \vec r_i  \; ,& 
\end{eqnarray}
where $M \equiv M_A + M_B = N_Am_A+N_Bm_B$ is the total mass, $m$ is
an arbitrary normalization mass, $\vec R_A$, $\vec R_B$ are the
individual center of mass coordinates and $\vec R$ is the overall
center-of-mass coordinate, i.e.
\begin{eqnarray}
\vec R = \frac{1}{M}\sum_{i} m_i \vec r_i  =
\frac{M_A\vec R_A+M_B\vec R_B}{M} \; .
\end{eqnarray}
The center of mass separation of the two components is then given by
the coordinate $\vec r \equiv \vec R_B - \vec R_A$.  The three length
coordinates, $\rho_A$, $\rho_B$, and $r$, are related to $\rho$ by
\begin{equation} \label{e25}
 m \rho^2 =  m_A \rho_A^2 +  m_B \rho_B^2 +  m_r r^2 \;,\;\; 
 m_r \equiv \frac{M_A M_B}{M_A + M_B} \;  .
\end{equation}
The remaining coordinates, beside $\rho_A$, $\rho_B$, $\vec R$, $\vec
r$, are all chosen as (hyper)angles denoted collectively for each
system by $\Omega_A$ and $\Omega_B$.

The total volume element is then given by
\begin{eqnarray}
\prod^{N_A}_{i=1}d^3r_i \prod^{N}_{i=N_A+1}d^3r_i
=N_A^{3/2}N_B^{3/2}\rho_A^{3N_A-4}\rho_B^{3N_B-4}  \nonumber \\ 
d^3R r^2 dr  d\rho_Ad\rho_B
d\Omega_{r}d\Omega_{N_A-1}d\Omega_{N_B-1} \; , \; \; \label{e30}
\end{eqnarray}
where $d\Omega_{r}$ is the angular volume element for $\vec r$ and
$d\Omega_{N_A-1},d\Omega_{N_B-1}$ accounts for the hyperangles
collected in $\Omega_A$ and $\Omega_B$, see \cite{sor02} for precise
definitions.

Using these coordinates we rewrite the external harmonic oscillator
potentials in the Hamiltonian (\ref{eq:TH}) as
\begin{eqnarray}
\frac{1}{2}m_A\omega^2\sum_{i=1}^{N_A}\vec r_{i}
+\frac{1}{2}m_B\omega^2\sum_{i=N_A+1}^{N}\vec r_{i} =   
\nonumber \\ \label{e35}
\frac{1}{2}M\omega^2R^2+\frac{1}{2}m_r\omega^2r^2
+\frac{1}{2}m_A\omega^2\rho_A^2+\frac{1}{2}m_B\omega^2\rho_B^2 \\  \nonumber
=\frac{1}{2}M\omega^2R^2+\frac{1}{2} m \omega^2\rho^2 \; .
\end{eqnarray}
Furthermore the kinetic energy operator $\hat T$ can also be separated
in terms related to total center of mass, relative and intrinsic $A$
and $B$ degrees of freedom, i.e.
\begin{eqnarray} \label{e40}
  \hat T = -\frac{\hbar^2}{2M}\vec \nabla_R^2  
    -\frac{\hbar^2}{2m_r}\vec \nabla_r^2 + \hat T_A+\hat T_B  \; ,
\end{eqnarray}
where $\nabla_R$ and $\nabla_r$ are the usual three-dimensional
derivatives with respect to the indicated coordinates, $\hat T_{A}$
and $\hat T_{B}$ are the intrinsic kinetic-energy operators expressed
by hyperspherical coordinates for each boson system, see \cite{sor02}.

The two-body interactions depend only on relative distances and we can
completely separate relative and total center of mass motions.
Furthermore, we can separate relative $A$ and $B$ except for the
coupling term $V_{AB}$.  The Hamiltonian in equation (\ref{eq:TH}) then
becomes
\begin{eqnarray}
\fl
H=H_{\rm c.m.}+H_{\rm rel}  \; ,\label{e45} \\
\fl
H_{\rm c.m.}=-\frac{\hbar^2}{2M}\vec \nabla_R^2
+\frac{1}{2}M\omega^2R^2 \; , \label{e50} \\
\fl
H_{\rm rel}=
T_{\rho_A}+T_{\rho_B}+T_{r}
+\frac{1}{2}m_A\omega^2\rho_A^2 
 +\frac{1}{2}m_B\omega^2\rho_B^2
+\frac{1}{2}m_r\omega^2r^2
+H_{\Omega} \; ,\label{e55} \\
\fl
H_{\Omega}=
\frac{\hbar^2}{2m_A}\frac{\hat \Lambda_{N_A-1}^2}{\rho_A^2}
+\frac{\hbar^2}{2m_B}\frac{\hat \Lambda_{N_B-1}^2}{\rho_B^2}
+\frac{\hbar^2}{2m_r}\frac{\hat L_r^2}{r^2}
+V_{A}+V_{B}+V_{AB} \; , \label{e60}\\
\fl
T_r=-\frac{\hbar^2}{2m_r}\frac{1}{r}\frac{\partial^2}{\partial r^2}r \;,\\
\fl
T_{\rho_A}=-\frac{\hbar^2}{2m_A} \rho_A^{2-3N_A/2}
\frac{\partial^2}{\partial \rho_A^2}  \rho_A^{3N_A/2-2} \;,\\
\fl
T_{\rho_B}=-\frac{\hbar^2}{2m_B} \rho_B^{2-3N_B/2}
\frac{\partial^2}{\partial \rho_B^2} \rho_B^{3N_B/2-2} \;,
\end{eqnarray}
where $\hat L_r^2$ is the usual angular momentum operator with
respect to $\vec r$, and $\hat \Lambda_{N_A-1}^2$ and $\hat
\Lambda_{N_B-1}^2$ consist of first and second order derivatives with
respect to the hyperangular degrees of freedom indicated by the
indices.  This Hamiltonian reduces for $m_A=m_B$ and identical
interactions, $g_{A} = g_{B} = g_{AB}$, to the one-component system
described in \cite{sor02}.

The separable center of mass motion is given by the harmonic
oscillator solutions to the Hamiltonian in equation (\ref{e50}). The
relative motion can be determined by the hyperspherical adiabatic
expansion method where the wave function for fixed values of
$(\rho_A,\rho_B,r)$ is expanded on the complete set of eigenfunctions
for $H_{\Omega}$ in equation (\ref{e60}). The leading term in the lowest
adiabatic channel is expected to consist of the lowest partial waves,
where the wave function is independent of all (hyper)angles. This is
the assumption valid in the large-distance limit where all particles
are far from each other and all directional dependence is averaged out
\cite{boh98,sor03}.  Then the wave function for the lowest eigenvalue
reduces to the form
\begin{equation} \label{e80}
 \Psi = \rho_A^{2-3N_A/2} \rho_B^{2-3N_B/2} r^{-1} f(\rho_A,\rho_B,r)  \; ,
\end{equation}
where $f$ only depends on the three radial coordinates.

This radial wave function and the corresponding eigenvalue $E_{\rm rel}$
are determined from the Schr\"odinger equation obtained by combining
equations (\ref{e55}) and (\ref{e80}), i.e.
\begin{eqnarray} 
\fl 
[-\frac{\hbar^2}{2m_A}\frac{\partial^2}{\partial \rho_A^2}
-\frac{\hbar^2}{2m_B}\frac{\partial^2}{\partial \rho_A^2}
-\frac{\hbar^2}{2m_r}\frac{\partial^2}{\partial r^2} \nonumber \\
\lo
+U_{\rm eff}(\rho_A,\rho_B,r)]f(\rho_A,\rho_B,r) 
=E_{\rm rel}f(\rho_A,\rho_B,r) \label{e85}  \\
\fl 
U_{\rm eff}(\rho_A,\rho_B,r)
=\frac{1}{2}\omega^2 \Big(m_A\rho_A^2
+\frac{1}{2}m_B\rho_B^2
+\frac{1}{2}m_rr^2\Big) \nonumber \\
\lo 
+ \frac{\hbar^2(3N_A-4)(3N_A-6)}{8m_A\rho_A^2}
+  \frac{\hbar^2(3N_B-4)(3N_B-6)}{8m_B\rho_B^2} \nonumber \\
\lo 
+\langle \Phi_0|V_{A}|\Phi_0 \rangle
+\langle \Phi_0|V_{B}|\Phi_0 \rangle
+\langle \Phi_0|V_{AB}|\Phi_0 \rangle  \; ,  \label{e90}
\end{eqnarray}
where $\langle \Phi_0|V|\Phi_0 \rangle$ is the expectation value of
the interaction $V$ with the constant angular wave function $\Phi_0$.
With the volume element in equation (\ref{e30}) we find by direct
integration
\begin{eqnarray}  \label{e95}
\langle \Phi_0|V_{A}|\Phi_0 \rangle
=&\frac{1}{\sqrt{2\pi}}\frac{\Gamma(\frac{3N_A-3}{2})}
{\Gamma(\frac{3N_A-6}{2})}N_A(N_A-1)\frac{\hbar^2 a_{A}}{m_A \rho_A^3} 
 \\ \label{e100}
\langle \Phi_0|V_{B}|\Phi_0 \rangle
=&\frac{1}{\sqrt{2\pi}}\frac{\Gamma(\frac{3N_B-3}{2})}
{\Gamma(\frac{3N_B-6}{2})}N_B(N_B-1)\frac{\hbar^2 a_{B}}{m_B \rho_B^3} \\
\label{eq:intAB}
\langle \Phi_0|V_{AB}|\Phi_0 \rangle
=& \frac{1}{\sqrt{2\pi}} 
\Big(\frac{\Gamma(\frac{3N_A-3}{2})}{\Gamma(\frac{3N_A-6}{2})}
\frac{\Gamma(\frac{3N_B-3}{2})}{\Gamma(\frac{3N_B-6}{2})}\Big)^{1/2} 
\frac{\hbar^2 a_{AB}}{2\mu_{AB} (\rho_A \rho_B)^{3/2} } \nonumber \\
&\times\sqrt{N_A(N_A-1)N_B(N_B-1)} 
I(\rho_A, \rho_B, r)\; ,
\end{eqnarray}
where $\Gamma(z)$ is the Gamma function and $I$ is derived in 
\ref{app:int} as a function of $\rho_A$, $\rho_B$ and $r$.

When $N_A \gg 1$ we have $\Gamma(\frac{3N_A-3}{2})
/\Gamma(\frac{3N_A-6}{2}) \approx (3N_A/2)^{3/2}$ and analogously for
$N_B \gg 1$, see \cite{boh98}.  Then $U_{rm eff}$ is given as
\begin{eqnarray}
\label{eq:Ueff}
U_{\rm eff} = \frac{1}{2}m_A\omega^2\rho_A^2
+\frac{1}{2}m_B\omega^2\rho_B^2
+\frac{1}{2}m_r\omega^2 r^2    
+\frac{9N_A^2\hbar^2}{8m_A\rho_A^2}
+\frac{9N_B^2\hbar^2}{8m_B\rho_B^2} \nonumber \\
+\frac{3^{3/2}\hbar^2}{2^2\pi^{1/2}} \Big(
\frac{N_A^{7/2}a_{A}}{m_A\rho_A^3}  
+ \frac{N_B^{7/2}a_{B}}{m_B\rho_B^3} 
+  \frac{ (N_A N_B)^{7/4} a_{AB}}
{2\mu_{AB}(\rho_A \rho_B)^{3/2} } I(\rho_A, \rho_B,r)\Big) \; ,
\end{eqnarray}
which can be considered as the energy surface depending on the three
collective coordinates $\rho_A$, $\rho_B$ and $r$.  The properties of
this effective potential then reflects the properties of the solutions
which, if necessary, could be computed from the eigenvalue
equation (\ref{e85}).

The expression reduces to the one-component potential
\cite{HME98,sor02} when $r=0$ and all particles are identical, i.e.
\begin{eqnarray} \label{e38}
 U_{\rm eff} = \frac{1}{2}m\omega^2\rho^2
+\frac{9N^2\hbar^2}{8m\rho^2} +
\frac{3^{3/2}\hbar^2}{2^2\pi^{1/2}} \frac{N^{7/2}a_s}{m \rho^3}  \; ,
\end{eqnarray}
where $a_s$ is the common scattering length and $m=m_A=m_B$.

\section{The effective potential}
\label{sec3}

To be specific we consider parameters corresponding to a system where
both components are $^{87}$Rb atoms with the mass $m_A= m_B= m= 1.44
\times 10^{-25}$kg \cite{HME98}.  The bosons in both components
are assumed to be trapped in the same spherical harmonic potential
with the frequency $\omega=2\pi\times23.5$Hz. The related oscillator
length $b_t=\sqrt{\hbar/m\omega}$ is then 2.22 $\mu$m.  The particle
numbers $N_A$, $N_B$ and the three $s$-wave scattering lengths
$a_{A}$, $a_{B}$ and $a_{AB}$ are variable parameters.

\subsection{The interaction $V_{AB}$}

\begin{figure}[hbt]
\vspace*{-0.0cm}
\begin{center}
\epsfig{file=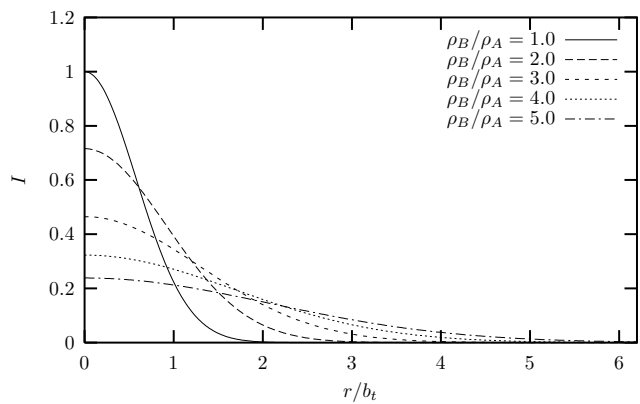,width=8.4cm,angle=0}
\end{center}
\vspace*{-0.5cm}
\caption{The interaction $I$ defined in equation (\ref{eq:intAB}) and 
\ref{app:int} as function of the distance $r$ between the two
centers of masses. The particle numbers are $N_A=N_B=20000$, $\rho_A =
100 b_t$ and the ratios $\rho_B /\rho_A$ are given on the figure. }
\label{fig1a}
\end{figure}

\begin{figure}[hbt]
\vspace*{-0.0cm}
\begin{center}
\epsfig{file=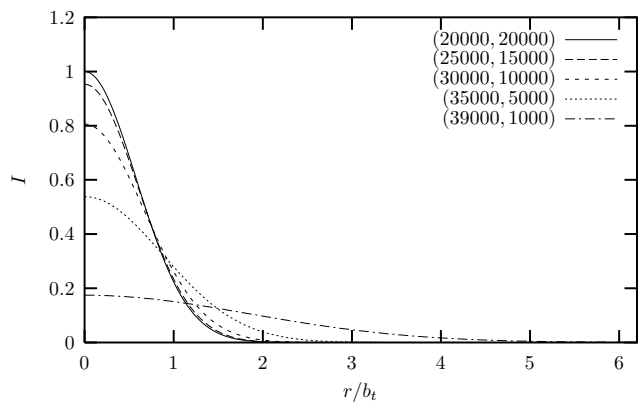,width=8.4cm,angle=0}
\end{center}
\vspace*{-0.5cm}
\caption{The interaction function $I$ defined in equation (\ref{eq:intAB}) 
and \ref{app:int} as function of the distance $r$ between the two
centers of masses for $\rho_A=\rho_B=100b_t$ and the values of $(N_A,
N_B)$ given on the figure. }
\label{fig1b}
\end{figure}

All terms in the potential (\ref{eq:Ueff}) are simple and
explicitly given, except the function $I$ arising from the interaction
$V_{AB}$ in equation (\ref{eq:intAB}), see appendix \ref{app:int}.  Without
this interaction the two subsystems would decouple and each behaves as
a separate one-component system.  We show this interaction term in
figures \ref{fig1a} and \ref{fig1b} for several sets of particle numbers
and hyperradii.

The general features are a monotonous function with a flat maximum at
$r=0$ and vanishing exactly for $r>\rho_A+\rho_B$. For given $N$ the
symmetric combination of $N_A=N_B$ and $\rho_A=\rho_B$ leads to the
highest maximum value of unity corresponding to the smallest average
distance between particles in the subsystems $A$ and $B$.  The
interaction between the two components is strongest for symmetric
systems with coinciding centers. As the centers are moved apart, the
interaction becomes larger for different values of $\rho_A$ and
$\rho_B$.  Asymmetric division of a given total number of particles
also decrease the interaction. These properties directly reflect the
dependences of the overlap between the two density distributions.

The two-body interaction strength $a_{AB}$ is a proportionality factor
on the function $I$. Consequently, attraction prefers symmetric
division and coinciding centers, whereas repulsion prefers precisely
opposite, i.e. asymmetry and large distance between the center of
masses.

\subsection{Dependence on center of mass distance}

We minimize the effective potential in equation (\ref{eq:Ueff}) with
respect to $\rho_{A}$ and $\rho_{B}$ for fixed $r$ and given
scattering lengths and particle numbers.  We first choose interactions
corresponding to the two-component condensate of two spin states of
$^{87}$Rb. The three scattering lengths are then almost identical and
given by $a_{A}=a_{B}=a_{AB}=103a_0$ where $a_0$ is the Bohr radius
\cite{HME98}.  The total number of particles is rather arbitrarily
fixed to be $N=N_A+N_B=40000$. The results for identical scattering
lengths are discussed in \cite{sog04}.

The main features are that the effective potential at $r=0$ is
independent of the division of particles $N_A$ and $N_B$, but still
strongly varying with the total number $N$. This potential value at
$r=0$ is the same as obtained for the one-component potential in
equation (\ref{e38}) when $\rho$ is calculated from equation (\ref{e25}) with
the minimum values of $\rho_{A}$ and $\rho_{B}$.

For these identical repulsive scattering lengths and the symmetric
divisions of $N_B = N_A$, the systems prefer coinciding centers of
mass, i.e. the potential has a minimum for $r=0$.  As we increase the
particle asymmetry, another minimum appears at $r \approx 2-3 b_t$ and
for $N_B < 0.134 N_A$ (or $N_A < 0.134 N_B$) eventually becomes
deeper. The minimum at $r=0$ remains, but now separated by a barrier.
The deepest of these minima is found when $N_B/N_A \approx 1/40$. Even
larger asymmetries again increase the depth and eventually only the
$r=0$ minimum remains as for a one-component system.

\begin{figure}[hbt]
\vspace*{-0mm}
\begin{center}
\epsfig{file=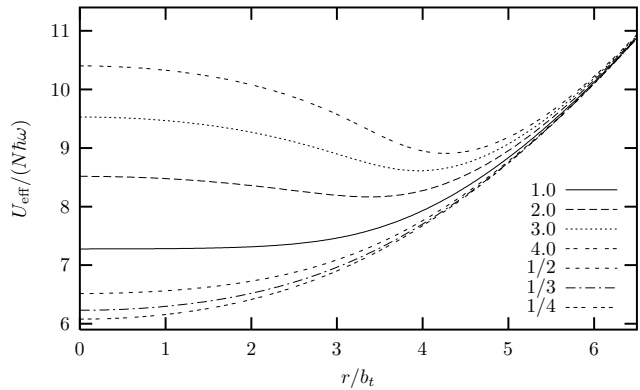,width=8.4cm,angle=0}
\end{center}
\vspace*{-0.5cm}
\caption[]{The effective potential in equation (\ref{e60}) as a function of 
$r/b_t$ for $^{87}$Rb masses $m_A = m_B = 1.44 \times 10^{-25}$kg,
trap length $b_t = 2.22 \mu$m ($\omega = 2\pi \times 23.5$ Hz),
particle numbers $N_A=N_B=20000$, and $a_A = a_B = 103 a_0$, where
$a_0$ is the Bohr radius. The curves correspond to the different
values of $a_{AB}/a_A$ given on the figure.}
\label{fig2}
\end{figure}

Maintaining $a_{A}=a_{B}>0$ we show in figure \ref{fig2} the variation
with the scattering length $a_{AB}>0$ for a symmetric division of
particle numbers. The flat region around $r=0$ is always present, but
the minimum for $a_{AB} \le a_{A}$ turns into a maximum when $a_{AB} >
a_{A}$. Then the repulsion between the subsystems is strong enough to
move the minimum at $r=0$ to finite values corresponding to values
$r/b_t \approx 3.5-4.5$. For less repulsion, $a_{AB} \le a_{A} =
a_{B}$, the minimum at $r=0$ is always present and in addition another
minimum at finite $r$ appears for sufficiently asymmetric particle
divisions.

The basic characteristics of a system are size and energy. The
root-mean-square distance is the measure of the size of a system which
in our coordinates are given by
\begin{eqnarray}  \label{e57}
 \bar r_A^2=\frac{1}{N_A}\langle \sum_{i=1}^{N_A}(\vec r_{i} -\vec R_A)^2 
 \rangle  =\frac{\langle \rho_A^2 \rangle}{N_A}  \; ,\\  \label{e58}
 \bar r_B^2=\frac{1}{N_B}\langle \sum_{i=N_A+1}^{N}(\vec r_{i} -\vec R_B)^2 
 \rangle  =\frac{\langle \rho_B^2 \rangle}{N_B} \; ,\\  \label{e59}
 \bar r^2=\frac{1}{M}\langle \sum_{i=1}^{N}m_i(\vec r_{i} -\vec R)^2 
 \rangle  =\frac{\langle m \rho^2 \rangle}{M}  \; ,
\end{eqnarray}
where $\bar r$, $\bar r_A$ and $\bar r_B$ are the root mean square
radii for total and individual subsystems and $\langle \rangle$
represents the expectation value for the total wave function.

For identical interactions of $a_{AB} = a_{A} = a_{B}$ we found in
\cite{sog04} that $\bar r_A = \bar r_B =  \bar r \approx 3 b_t$  in the 
minimum at $r=0$ independent of particle division.  This structure
closely resembles the one-component system.  Increasing the distance
$r$ between the centers of the subsystems reduce all three root mean
square radii.  However, as $N_B/N_A$ changes, the system with the
largest particle number always maintains roughly the same value of
about $3b_t$. In contrast, the root mean square radius of the other
subsystem decreases to about half of the value for $r=0$.

The minima of effective potentials as in figure \ref{fig2} define the
preferred structures.  The effect of the repulsion is that the two
systems try to avoid each other while staying at distances small
compared to the size of the confining external field. The large
subsystem is exploiting the space almost like it was alone in the
trap. When one subsystem is small, it becomes advantageous to place
about half of its particles outside the other subsystem. Then the
repulsion on the small subsystem from the trap and the other subsystem
is minimized.

\begin{figure}[hbt]
\vspace*{-0mm}
\begin{center}
\epsfig{file=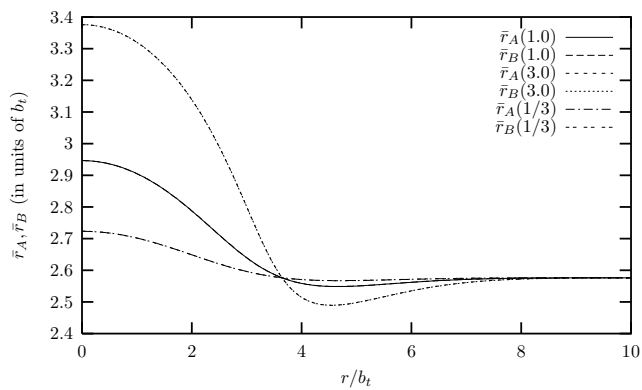,width=8.4cm,angle=0}
\end{center}
\vspace*{-0.5cm}
\caption[]{The root-mean-square radii $\bar r_A = \bar r_B$ as 
a function of $r$ for some of the systems in figure \ref{fig2}.  The
curves again correspond to the different values of $a_{AB}/a_A$ given
on the figure. }
\label{fig3}
\end{figure}

When $N_B < 0.134 N_A$, the lowest minimum occurs at finite $r$.  The
center of mass of the small system $B$ remains within the radius $\bar
r_A$ until $N_B \approx 0.004 N_A$. For even smaller subsystems the
center of $B$ moves a little outside $\bar r_A$, but still the
distance between the centers at the minimum remains smaller than $\bar
r_A + \bar r_B$. Therefore, complete separation between the systems
does not occur.

Many of these basic properties remain for $0 < a_{AB} \neq a_{A} =
a_{B}$ as shown in figure \ref{fig3}. The root mean square radii again
are largest for $r=0$ with values increasing as a function of $a_{AB}$
corresponding to decreasing the overlap in order to decrease the energy.  For
relatively large values of $r$ where the subsystems essentially can
avoid overlapping, the sizes become independent of $a_{AB}$. At
intermediate distances a minimum size is always found.  The smaller
the repulsion $a_{AB}$, the less is the variation with center of mass
distance.

\subsection{Curvature and normal modes}

The local minima in the $r$-direction represent a projection in the
three dimensional space. We can evaluate the stability by computing
the second derivatives of the effective potential in these points. The
kinetic energy operators already only contain second derivatives and
the relative Hamiltonian $H_{\rm rel}$ is then to second order
approximated by
\begin{eqnarray}
H_{\rm rel}\approx   U_{\rm eff}(\rho_{A \rm min},\rho_{B \rm min},r_{\rm min})
 \nonumber \\ 
 -\frac{\hbar^2}{2m_A}\frac{\partial^2}{\partial \rho_A^2}
-\frac{\hbar^2}{2m_B}\frac{\partial^2}{\partial \rho_B^2}
-\frac{\hbar^2}{2m_r}\frac{\partial^2}{\partial r^2} \nonumber \\
 +\frac{1}{2}
\left(\begin{array}{ccc}
\rho_A-\rho_{A \rm min}, & \rho_B-\rho_{B \rm min}, & r-r_{\rm min}
\end{array}\right) \nonumber \\ 
\left(\begin{array}{ccc}
\left.\frac{\partial^2U_{\rm eff}}{\partial \rho_A^2}\right|_{\rm min}&
\left.\frac{\partial^2U_{\rm eff}}{\partial \rho_A \partial \rho_B}\right|_{\rm min}&
\left.\frac{\partial^2U_{\rm eff}}{\partial \rho_A \partial r}\right|_{\rm min}\\
\left.\frac{\partial^2U_{\rm eff}}{\partial \rho_B \partial \rho_A}\right|_{\rm min}&
\left.\frac{\partial^2U_{\rm eff}}{\partial \rho_B^2}\right|_{\rm min}&
\left.\frac{\partial^2U_{\rm eff}}{\partial \rho_B \partial r}\right|_{\rm min} \\
\left.\frac{\partial^2U_{\rm eff}}{\partial r \partial \rho_A}\right|_{\rm min}&
\left.\frac{\partial^2U_{\rm eff}}{\partial r \partial \rho_B}\right|_{\rm min}&
\left.\frac{\partial^2U_{\rm eff}}{\partial r^2}\right|_{\rm min}
\end{array}
\right) 
\left(
\begin{array}{c}
\rho_A-\rho_{A \rm min} \\
\rho_B-\rho_{B \rm min} \\
r-r_{\rm min}
\end{array}
\right)
\label{eq:qf}
\end{eqnarray}
where we used the notation of the $3 \times 3$-matrix sandwiched
between the two vectors. All first order derivatives are zero in these
minima.

The approximation in equation (\ref{eq:qf}) is almost directly three
separable harmonic oscillators. We first redefine the variables by
scaling with the related masses as
\begin{eqnarray} \label{e135}
\delta \rho_A &\equiv& (\rho_A-\rho_{A \rm min}) \sqrt{m_A\omega/\hbar}\;,
 \\ \label{e137}
\delta \rho_B &\equiv& (\rho_B-\rho_{B \rm min}) \sqrt{m_B\omega/\hbar}\;, \\
\delta r &\equiv& (r-r_{\rm min}) \sqrt{m_r\omega/\hbar} \; . \label{e139} 
\end{eqnarray}
Then all three kinetic energy operators are dimensionless and the
energy unit is $\hbar\omega$. It is now sufficient to diagonalize
the potential matrix in equation (\ref{eq:qf}).  The total Hamiltonian then
has eigenvalues and eigenvectors $\lambda_i $ and $\vec
v_i=(v^A_i,v^B_i,v^r_i)$ for $i=1,2,3$. The total energy is given by
$E=\hbar\omega\sum_i^3 (n_i+1/2)\sqrt{\lambda_i}$ with the
non-negative integers $n_i$ corresponding to one-dimensional harmonic
oscillators. The excitation energies for each of these vibrational
degrees of freedom are then integer multiples of $\hbar\omega
\sqrt{\lambda_i}$.

For the minimum at $r=0$ the $\rho_A$-$\rho_B$ and the $r$ degrees of
freedom decouple completely.  This is a consequence of the flat
maximum at $r=0$ for the interaction function in figures \ref{fig1a} and
\ref{fig1b}, because first order $r$-derivatives of $I$ then vanish and 
the only other $r$-dependence is the $r^2$ arising from the external
field.  For this minimum the lowest vibrational $r$-mode carries the
energy $\hbar \omega (n_r+3/2) \sqrt{\lambda_r}$ dictated by the
boundary condition at $r=0$ corresponding to negative parity. The
total energy is in this case given by $E=\hbar\omega (
(n_3+3/2)\sqrt{\lambda_3} + \sum_{i=1}^2(n_i+1/2)\sqrt{\lambda_i})$.
Still the vibrational excitation energies are integer multiples of
$\hbar\omega \sqrt{\lambda_i}$.

The normal modes are given by the components of the eigenvectors $\vec
v_i$ defining a direction in the coordinate system $(\delta \rho_A,
\delta \rho_B, \delta r)$.  The physical system has natural length 
scales for the different degrees of freedom corresponding to the
initial coordinates $(\rho_A,\rho_B,r)$. They are exhibited by the
sizes of the subsystems, i.e. the root mean square radii defined in
equations (\ref{e57})-(\ref{e59}).  Thus the proper physical length scales
are $\rho_A/\sqrt{N_A}$ and $\rho_B/\sqrt{N_B}$, which means that we
should multiply the coefficient measuring the $\rho_A$-component by
$\sqrt{N_A}$ to change to the coefficient measuring the corresponding
root mean square radius.  The normal mode directions with these root
mean square unit vectors are then represented by the vectors $(\tilde
v^A_i, \tilde v^B_i, \tilde v^r_i) \equiv (v^A_i
\sqrt{N_Am_A\omega/\hbar}, v^B_i \sqrt{N_Bm_B\omega/\hbar}, v^r_i
\sqrt{m_r\omega/\hbar})$.

These eigenvectors and the related eigenvalues are given in tables
\ref{tab1} and \ref{tab2}.   For the minimum at $r=0$ the $r$-direction 
is completely decoupled from the two other normal modes mixing
$\rho_A$ and $\rho_B$. By far the lowest eigenvalue corresponds to the
$r$-mode which means vibrations of the relative center of mass.

\begin{table}
\caption{\label{tab1}
Eigenvalues and eigenvectors for the second order harmonic 
motion in the minimum at $r=0$ of the effective potential. The
eigenvectors ($\tilde v^A_i, \tilde v^B_i \tilde v^r_i$) refer to the
units of root mean square radii or in terms of the initial coordinates
($(\rho_A-\rho_{A \rm min})\sqrt{N_Am_A\omega/\hbar}, (\rho_B-\rho_{B \rm min})
\sqrt{N_Bm_B\omega/\hbar}, (r-r_{\rm min}) \sqrt{m_r\omega/\hbar}$).
The scattering lengths are $a_{AB} = a_{A} = a_{B} = 103 a_0$.}
\begin{tabular}{cc||cccc|cccc|cccc}
\br
$N_A$ & $N_B$ &
$\lambda_1$ & $\tilde v^A_1$ & $\tilde v^B_1$ & $\tilde v^r_1$ &
$\lambda_2$ & $\tilde v^A_2$ & $\tilde v^B_2$ & $\tilde v^r_2$ &
$\lambda_3$ & $\tilde v^A_3$ & $\tilde v^B_3$ & $\tilde v^r_3$ \\
\mr
20000 & 20000 &
4.97 & 0.71 & 0.71 & 0.0 &
2.54 & 0.71 & -0.71 & 0.0 & 
0.03 & 0.0 & 0.0 & 1.0 \\
32000 & 8000 &
4.97 & 0.97 & 0.24 & 0.0 &
2.54& 0.71 & -0.71 & 0.0 &
0.03 & 0.0 & 0.0 & 1.0 \\
36000 & 4000 &
4.97 & 0.99 & 0.11 & 0.0 &
2.55 & 0.71 & -0.71 & 0.0 &
0.03 & 0.0 & 0.0 & 1.0 \\
37000 & 3000&
4.97 & 1.00 & 0.08 & 0.0 &
2.55 & 0.71 & -0.71 & 0.0 &
0.03 & 0.0 & 0.0 & 1.0 \\
39000 & 1000 &
4.97 & 1.00 & 0.03 & 0.0 &
2.55 & 0.71 & -0.71 &  0.0 &
0.03 & 0.0 & 0.0 & 1.0 \\
39800&200&
4.97 & 1.00 & 0.01 & 0.0 &
2.56 & 0.71 & -0.71 & 0.0 &
0.04 & 0.0 & 0.0 & 1.0 \\
\br
\end{tabular}
\end{table}

\begin{table}
\caption{\label{tab2}
Eigenvalues and eigenvectors for the second order harmonic 
motion in the minimum at $r \neq0$ of the effective potential. The
eigenvectors ($\tilde v^A_i, \tilde v^B_i \tilde v^r_i$) refer to the
units of root mean square radii or in terms of the initial coordinates
($(\rho_A-\rho_{A \rm min})\sqrt{N_Am_A\omega/\hbar}, (\rho_B-\rho_{B \rm min})
\sqrt{N_Bm_B\omega/\hbar}, (r-r_{\rm min}) \sqrt{m_r\omega/\hbar}$).
The scattering lengths are $a_{AB} = a_{A} = a_{B} = 103 a_0$.}
\begin{tabular}{cc||cccc|cccc|cccc}
\br
$N_A$ & $N_B$ &
$\lambda_1$ & $\tilde v^A_1$ & $\tilde v^B_1$ & $\tilde v^r_1$ &
$\lambda_2$ & $\tilde v^A_2$ & $\tilde v^B_2$ & $\tilde v^r_2$ &
$\lambda_3$ & $\tilde v^A_3$ & $\tilde v^B_3$ & $\tilde v^r_3$ \\
\mr 
20000 & 20000 &
-&-&-&-&-&-&-&-&-&-&-&- \\
32000 & 8000 &
-&-&-&-&-&-&-&-&-&-&-&- \\
36000 & 4000 &
4.97 & 0.99 & 0.08 & 0.08 &
3.29 & 0.72 & -0.62 & -0.31 &
0.26 & 0.26 & 0.49 & -0.83 \\
37000 & 3000&
4.97 & 1.0 & 0.06 & 0.06 &
3.47 & 0.71 & -0.62 & -0.34 &
0.48 & 0.32 & 0.51 & -0.80 \\
39000 & 1000 &
4.97 & 1.0 & 0.01  & 0.02 &
3.81 & 0.67 & -0.63 & -0.40 &
1.21 & 0.45 & 0.50 & -0.75 \\
39800&200&
4.97 & 1.0 & 0.0 & 0.0 &
3.90 & 0.60 & -0.67 & -0.44 &
1.77 & 0.53 & 0.47 & -0.71 \\
\br
\end{tabular}
\end{table}

The next mode is remarkably independent of the particle division,
always with exactly equal amplitudes of opposite phase in the $A$ and
$B$ root mean square radii. This mode is then an isovector breathing
mode where one subsystem shrinks radially while the other radially
expands. This vibration is around the equilibrium structure where
the subsystems both are of equal size.

The highest-lying mode for the $r=0$ minimum corresponds to in-phase
oscillations of $\rho_A$ and $\rho_B$. The equal amplitudes of the
real size components, $(1,-1) \propto (\tilde v^A_2,\tilde v^B_2)
\propto (v^A_2\sqrt{N_A}, v^B_2 \sqrt{N_B})$, combined with
orthogonality of $(v^A_1,v^B_1)$ and $(v^A_2,v^B_2) \propto
(\sqrt{N_B},- \sqrt{N_A}) $, implies that $(v^A_1,v^B_1) \propto
(\sqrt{N_A}, \sqrt{N_B})$, and $(\tilde v^A_1,\tilde v^B_1) \propto
(N_A,N_B) $.

In general, the vector $(v^A, v^B, v^r) \propto (\rho_{A \rm min}
\sqrt{m_A\omega/\hbar}, \rho_{B \rm min}\sqrt{m_B\omega/\hbar}, 
r_{\rm min} \sqrt{m_r\omega/\hbar})$ is at the minimum position precisely
in the direction of the total hyperradius defined in equation (\ref{e25}).
For $r=0$ the root mean square radii of the two subsystems are equal
and we have from equations (\ref{e57})-(\ref{e59}) that 
$\rho_{A \rm min} \sqrt{N_B} \approx \rho_{B \rm min} \sqrt{N_A}$.
Thus the highest-lying vibrational mode, $(v^A_1,v^B_1) \propto
(\sqrt{N_A}, \sqrt{N_B})$, is exactly in the direction of the
hyperradius.

For an equal particle division this is fully the physical breathing mode
where both subsystems move in phase with an equal scaling of their
radii. For asymmetric particle divisions the largest subsystem
tries to breathe like it was alone. The smallest then has to follow as
well as possible consistent with orthogonality, which in turn is
equivalent to following the $\rho$-direction. Equal breathing amplitudes
are then only consistent with the $\rho$-direction for a symmetric
particle division.

For the minimum at $r\neq0$ all three directions are mixed in the
normal modes as seen in table \ref{tab2}.  The lowest-lying energy is
now much larger than for the $r=0$ minimum, but the largest
probability is still related to vibration in the $r$-direction.  The
second mode again corresponds almost completely to the isovector
breathing mode, i.e. out-of-phase oscillation with equal real size
amplitudes of the two subsystems. Now an admixture of up to 30\%
probability in the $r$-direction is present. 

The highest-lying mode is in-phase oscillation of the two subsystems.
Again this mode corresponds to maximum variation of the total
hyperradius and therefore the breathing mode cannot be fully
exploited.  The reason is that the largest subsystem determines the
mode and tries to oscillate as it was alone in the trap.

For both minima the smallest eigenvalue corresponds essentially to the
$r$-direction. This mode exploits that only the repulsion between $A$
and $B$ is changing when the subsystems maintain their sizes and only
move their centers of mass. The second and third eigenvalues are
essentially the isovector and isoscalar breathing modes, respectively.
The highest vibrational excitation energy,
$\hbar\omega\sqrt{\lambda}$, is consistently about
$\hbar\omega\sqrt{5}$ which is the one-component result for a
Bose-Einstein condensate in the limit of large particle numbers
\cite{pit03}. This breathing mode is energetically less favored than
the isovector breathing mode, where one subsystem oscillates against
the other and the repulsive interaction minimizes the overlap. In both
these modes all three repulsions contribute.

\section{Stability conditions}
\label{sec4}

The stability is first of all determined by the existence of minima.
However, for many-body atomic systems lower lying minima certainly
exist with structures completely different from condensed states.
These metastable states are confined by barriers providing a finite
lifetime.  In three-dimensional quantum mechanics the minimum in the
potential energy may be too shallow or narrow to allow the zero point
oscillations.  Crude estimates of stability are then first obtained by
locating extremum points with positive curvature in the potential.
Second by comparing the vibrational energies with the barrier heights.

\subsection{The $r$-mode for identical interactions}

\begin{figure}[hbt]
\vspace*{-0.0cm}
\begin{center}
\epsfig{file=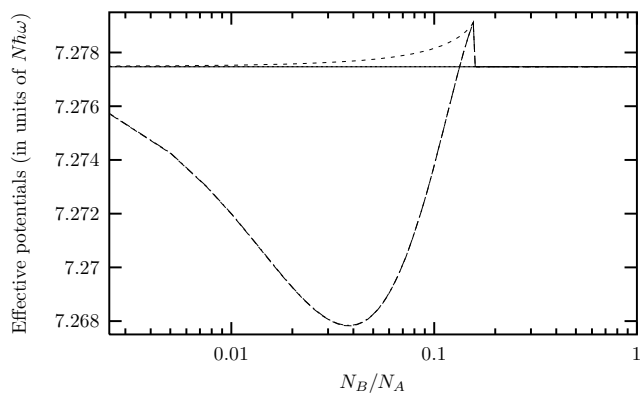,width=8.4cm,angle=0}
\end{center}
\vspace*{-0.5cm}
\caption[]{The values of the effective potential at the extremum points 
as a function of $N_B/N_A$ for $N_A + N_B = 40000$ and $a_{A} = a_{B} =
a_{AB} = 103 a_0$.  We show the minima at $r=0$ (solid line) and
$r\neq0$ (long-dashed line), and the separating maximum (short-dashed
line).  The zero point energies, added to the effective potentials,
are also shown for both minima, $r=0$ (dotted line) and $r\neq0$
(dot-dashed line).}
\label{fig4}
\end{figure}

Let us first assume identical scattering lengths where two minima in
the $r$-direction appears for some divisions of particles $N_A$ and
$N_B$. The values of the effective potential at these two minima for
$r=0$ and $r\neq0$ are shown in figure \ref{fig4}. The first prominent
feature is that only the $r=0$-minimum exists for $N_B/N_A > 5400/34600
\sim 0.156$. For more asymmetric divisions, $N_B/N_A < 4712/35288 \sim 
0.134$, the minimum at finite $r$ quickly becomes substantially
deeper.

The zero point energies are added to the potential, but the sum
coincides with the potential within the thickness of the lines in
figure \ref{fig4}. Both oscillations are essentially always classically
allowed, i.e. around finite $r$ very quickly after it appeared and
around $r=0$ until the asymmetry reaches very small values.

\begin{figure}[hbt]
\vspace*{-0mm}
\begin{center}
\epsfig{file=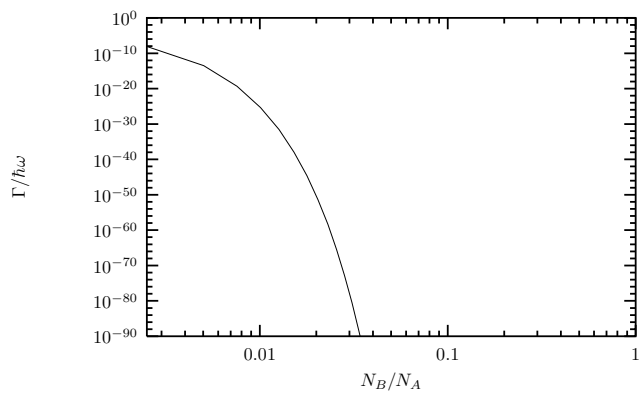,width=8.4cm,angle=0}
\end{center}
\vspace*{-0.5cm}
\caption[]{The WKB decay rate defined in equation (\ref{e160})
as function of $N_B/N_A$ for the minimum at $r=0$ for the cases in 
figure \ref{fig4}. }
\label{fig10}
\end{figure}

The highest minimum located at $r=0$ could in principle decay into the
lowest minimum at finite $r$. The related lifetime can be estimated by
the WKB tunneling approximation where the decay rate $\Gamma/\hbar$ is
given by
\begin{eqnarray} \label{e160}
\fl
  \Gamma/\hbar = \frac{\omega_r}{\pi} \exp\Big(-\frac{2}{\hbar} 
 \int^{r_1}_{r_2} {\rm d} r 
  \sqrt{2 m_r(U_{\rm eff}(r) - U_{\rm eff}(r=0) -  
 3 \hbar \omega_r/2)}\Big) \; ,
\end{eqnarray}
where $r_1$ and $r_2$ are the turning points and $\omega_r = \omega
\sqrt{\lambda_r}$ is the  oscillation frequency in the
$r$-direction.  These decay rates are computed and shown in
figure \ref{fig10} for the $r=0$ minimum for the cases exhibited in
figure \ref{fig4}.  The lifetime increases dramatically as the systems
become more asymmetric.  Only when $N_B/N_A$ is close to 0.001, the rate
becomes comparable to the frequency of the trap. For example for
$N_B/N_A \approx 0.0025$ the rate $\Gamma/(\hbar \omega)\approx 8
\times 10^{-9}$.

\begin{figure}[hbt]
\vspace*{-0.0cm}
\begin{center}
\epsfig{file=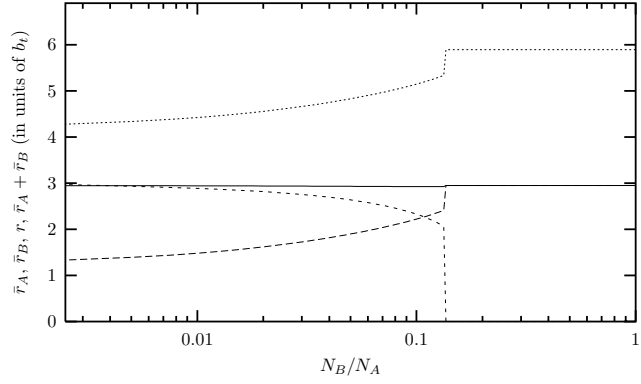,width=8.4cm,angle=0}
\end{center}
\vspace*{-0.5cm}
\caption[]{The root mean square radii $\bar r_A$ (solid), $\bar r_B$ 
(long-dashed), $r$ (short-dashed) and $\bar r_A+\bar r_B$ (dotted) for
the minimum at finite $r$ shown in figure \ref{fig4}. The minimum at
$r=0$ has $\bar r_A = \bar r_B \approx 3 b_t$ which also is shown for
the most symmetric systems.}
\label{fig5}
\end{figure}

The structures in these minima are reflected by the root mean square
radii defined in equations (\ref{e57})-(\ref{e59}). For $r=0$ all these sizes
are equal. For finite $r$ we show the results in figure \ref{fig5} for
the same parameters as in figure \ref{fig4}.  The largest system
maintains the same size $\bar r_A \approx 3 b_t$ independent of
particle number. The size of the smallest system decreases with
increasing asymmetry. In all cases we find that $\bar r_A+\bar r_B >
r$, i.e. complete separation between systems never occurs.
Furthermore, $\bar r_A > \bar r$ when $N_B/N_A > 150/39850 \sim
0.004$, i.e. the center of mass of the small system remains within the
radius of the large system except for very asymmetric particle
divisions.

\subsection{The $r$-mode for symmetric particle division}

The stability of the $r$-mode changes with the three interactions.
For $r=0$ the condition for instability against separation of the two
center of masses is a negative curvature in the minimum of the
effective potential in the $r$-direction.  We restrict ourselves to
$m_A=m_B=m$, $N_A=N_B$ and $a_A=a_B$ and define the coordinate at the
potential minimum by $\rho_{A \rm min}=\rho_{B \rm min}$.  
This minimum point is
determined as the zero point of the first derivative,
i.e. $\frac{\partial U_{\rm eff}}{\partial \rho_A} |_{r=0}=0$, which is
equivalent to
\begin{eqnarray}
\label{eq:sepcon2}
  f(\rho_A) \equiv [m\omega^2\rho_A^5-\frac{9}{4}
 \frac{N_A^2\hbar^2}{m}\rho_A  
-\frac{3^{5/2}}{4\sqrt{\pi}}\frac{N_A^{7/2}\hbar^2}{m}(a_{A}+a_{AB})] = 0\;,
\end{eqnarray}
with precisely one solution $\rho_{A \rm min}/b_t \geq
(3N_A/2)^{1/2} 5^{-1/4}$ when $N_A(a_{A}+a_{AB})/b_t
\geq - (8\pi)^{1/2} 5^{-5/4}$.  Furthermore, the function
$f(\rho_A)$ is positive for $\rho_A>\rho_{A \rm min}$.

A negative curvature, $\frac{\partial^2 U_{\rm eff}}{\partial
r^2}|_{r=0}<0$, is then equivalent to
\begin{eqnarray} \label{eq:sepcon1}
\rho_A^5 < \frac{3^{5/2}N_A^{7/2}}{2\pi^{1/2}}
\frac{\hbar^2 a_{AB}}{m^2\omega^2} \equiv \rho_{Ac}^5
\end{eqnarray}
The instability condition, $(3N_A/2)^{1/2} 5^{-1/4} b_t \leq
\rho_{A \rm min} < \rho_{Ac}$, is equivalent to the inequality
$f(\rho_{Ac})>0$, which in turn is equivalent to the two simultaneous
conditions
\begin{eqnarray}
\label{e171}
\frac{a_{A}}{b_t} &<&
\frac{a_{AB}}{b_t}
-\frac{\pi^{2/5}}{2^{1/5}N_A^{4/5}}
\left(\frac{a_{AB}}{b_t}\right)^{1/5} \; , \\  \label{e165}
\frac{N_Aa_{AB}}{b_t} &>& \frac{\pi^{1/2}}{2^{3/2}5^{5/4}} \; .
\end{eqnarray}

When $a_{A}=a_{B}=0$ the critical repulsion derived from
equation (\ref{e165}) corresponds to
$a_{AB}/b_t=\frac{\sqrt{\pi}}{2^{1/4}N_A}$. For finite values of
$a_{A}=a_{B}>0$ the critical value of $a_{AB}$ has to be
correspondingly increased.  The behavior of the effective potential is
illustrated in figure \ref{fig2} for parameter combinations close the
critical values for separation of the two centers of mass.  The
potentials are rather flat and the minimum therefore quickly moves to
separation distances comparable to the trap length.

\subsection{Stability of the $\rho_A$ and $\rho_B$-modes}

Qualitative understanding can be reached by analytic investigations
for symmetric systems, i.e. $N_A=N_B$, $a_{A}=a_{B}$ with the
constraint $r=0$. Then the condition that the effective potential at
$r=0$ has no stationary point, $\frac{\partial U_{\rm eff}}{\partial
\rho_{A}}|_{r=0}\neq0$  for any coordinates $\rho_A$ and
$\rho_B$, can be expressed as
\begin{equation}
\label{eq:stationary}
\frac{N_A(a_{A}+a_{AB})}{b_t}
\leq - \frac{2^{3/2}\pi^{1/2}}{5^{5/4}} \sim -0.67  \; .
\end{equation}
Then the two-component system is unstable since collapse to a point is
unavoidable.

Stability requires, beside a stationary point, also that the curvature
is positive in all directions. Then the eigenvalues of the curvature
matrix must all be positive at the stationary point.  For $r=0$ we
then construct and diagonalize the matrix
\begin{eqnarray}
\label{eq:matrix}
\left(\begin{array}{cc}
\frac{\partial^2 U_{\rm eff}}{\partial \rho_A^2} &
\frac{\partial^2 U_{\rm eff}}{\partial \rho_A \partial \rho_B} \\
\frac{\partial^2 U_{\rm eff}}{\partial \rho_B \partial \rho_A} &
\frac{\partial^2 U_{\rm eff}}{\partial \rho_B^2} 
\end{array}\right) \; .
\end{eqnarray}
For $\rho_A=\rho_B$ we can compute the eigenvalues analytically and
find the conditions expressing when at least one of them is negative
and the point therefore unstable. In \ref{app:int2} we derive
these conditions, i.e. the system is unstable when the following two
inequalities both are satisfied
\begin{eqnarray}
\label{eq:stable-rho}
\frac{a_{AB}}{b_t} &\leq& 
-\frac{5^5N_A^4}{2^8\pi^2}
\left(\frac{a_{A}}{b_t}\right)^5
+\frac{1}{4}\frac{a_{A}}{b_t}  \quad , \\ \label{e173}
\frac{N_Aa_{A}}{b_t} &\leq& - \frac{2^{3/2}\pi^{1/2}}{5^{5/4}} \; .
\end{eqnarray}
The last inequality, the stability condition for the $A$-component
alone, can be obtained both in the mean-field approximation and with
hyperspherical coordinates \cite{pet01,pit03,boh98}. Furthermore,
equation (\ref{e173}) is obtained from equation (\ref{eq:stable-rho}) for
non-interacting subsystems, where $a_{AB}=0$.  Both
equations (\ref{eq:stationary}) and (\ref{eq:stable-rho}) can be derived in
the mean field approximation \cite{BCP97}.

\begin{figure}[hbt]
\vspace*{-0mm}
\begin{center}
\epsfig{file=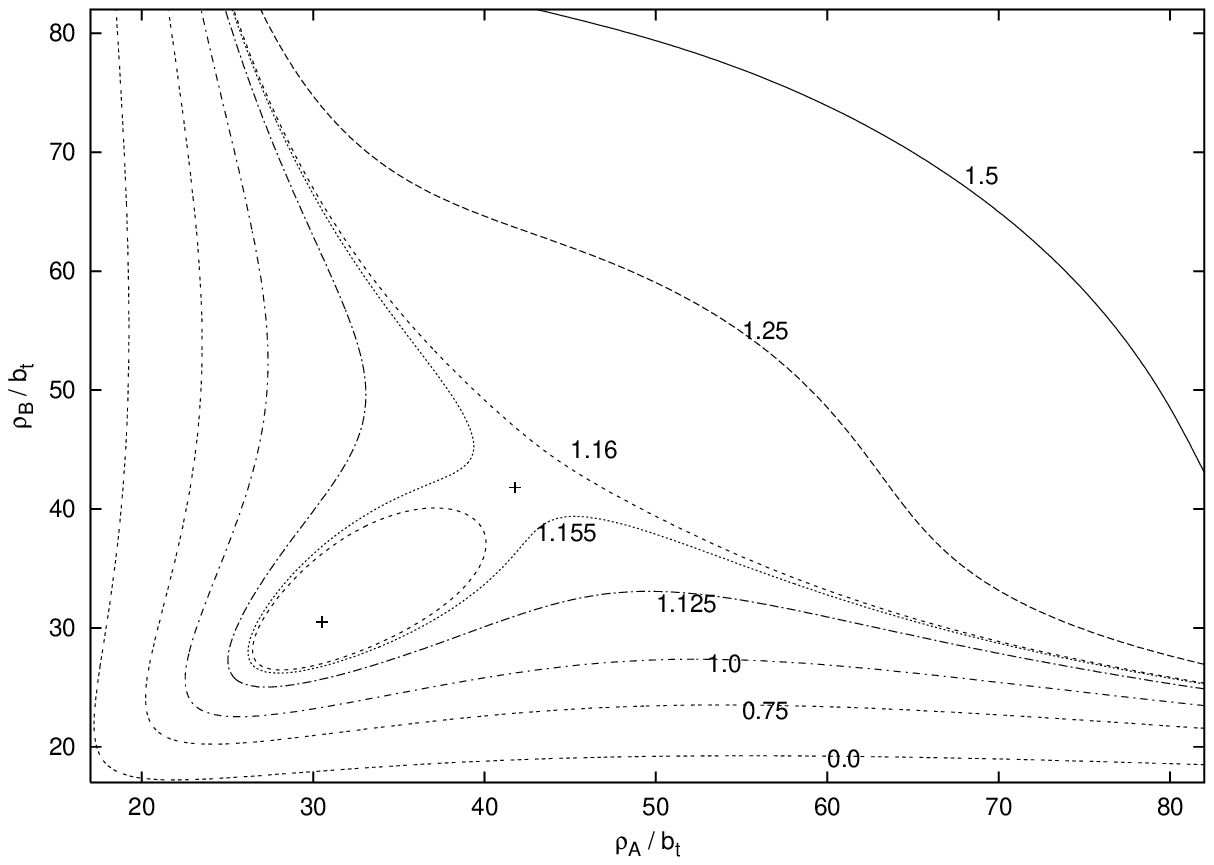,width=8.8cm,angle=0}
\end{center}
\vspace*{-0.5cm}
\caption[]{Contour plot of the potential $U_{\rm eff}/(N\hbar\omega)$
in equation (\ref{eq:Ueff}) as a function of $\rho_A$ and $\rho_B$ for $r=0$,
$N_A = N_B = N/2 =2000$, $a_{AB}=0.0001 b_t$, and
$a_{A}=a_{B}=-0.00042 b_t$. }
\label{fig9}
\end{figure}

These conditions are most easily visualized by two-dimensional contour
diagrams of the effective potential as function of $\rho_A$ and
$\rho_B$ with $r=0$. Such a diagram is shown in figure \ref{fig9} for
interactions very close to the critical values given in
equations (\ref{eq:stable-rho}) and (\ref{e173}).  The potential increases
when either $\rho_A$ or $\rho_B$ individually become large. This
effect of the external field is especially clearly seen along the line
$\rho_A = \rho_B$, where the potential decreases from large values at
very large distances, proceeds through a minimum located at around $42
b_t$ to reach a maxium at about $30 b_t$, and finally decreases to
$-\infty$ at the origin.  However, the minimum along this line is in
fact a saddle point and therefore unstable as soon as both degrees of
freedom are included. From the saddle point the energy decreases in
the ``isovector'' direction where the sum of $\rho_A$ and $\rho_B$
remains constant.

\begin{figure}[hbt]
\vspace*{-0mm}
\begin{center}
\epsfig{file=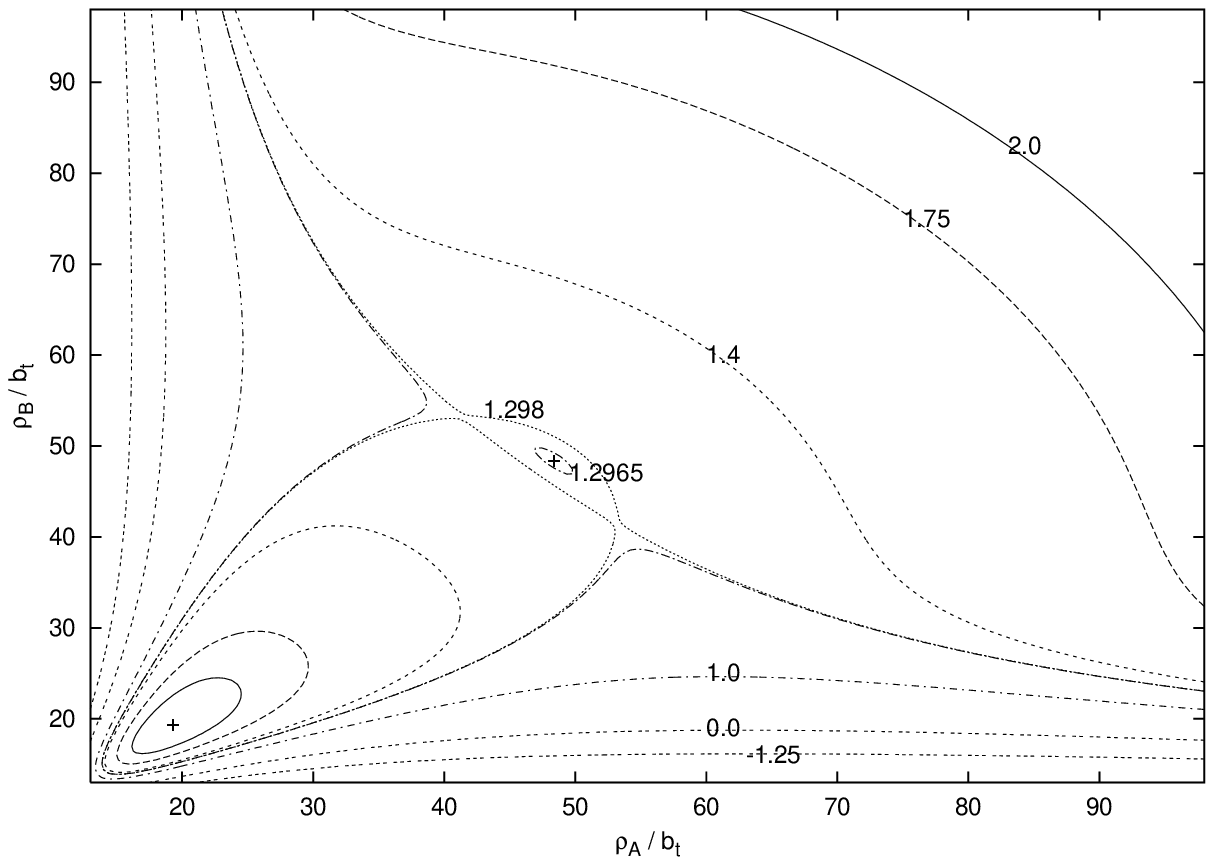,width=8.8cm,angle=0}
\end{center}
\vspace*{-0.5cm}
\caption[]{Contour plot of the potential $U_{\rm eff}/(N\hbar\omega)$
in equation (\ref{eq:Ueff}) as a function of $\rho_A$ and $\rho_B$ for $r=0$,
$N_A = N_B = N/2 =2000$, $a_{AB}=0.0002 b_t$, and
$a_{A}=a_{B}=-0.00042 b_t$. }
\label{fig8}
\end{figure}

This saddle point changes into a stable minimum with slightly larger
repulsion between the two subsystems as illustrated in
figure \ref{fig8}.  The same qualitative behavior is seen along the
$\rho_A = \rho_B$ line, where the minimum and maximum are shifted to
positions further apart at about $48 b_t$ and $20 b_t$, respectively.
Also the regions at relatively large distances along the coordinate
axes are essentially unchanged. In the ``isovector'' direction, as
well as in all other directions, the potential now increases as
characteristic for a minimum.  However, the minimum is very shallow
and only a small amount of  energy would destabilize the system.

\subsection{Phase diagram of stability}

\begin{figure}[hbt]
\vspace*{-0.0cm}
\begin{center}
\epsfig{file=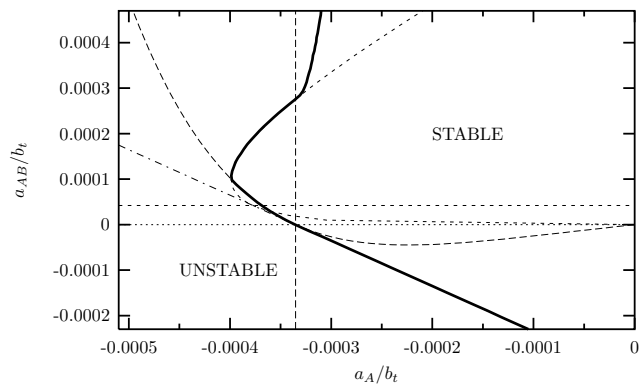,width=8.4cm,angle=0}
\end{center}
\vspace*{-0.5cm}
\caption[]{Stability regions for a two-component system with 
$N_B = N_A = 2000$ as a function of $a_A = a_B < 0$ and $a_{AB}$. The
thick solid line separates stable (right side) from unstable (left
side) regions when we allow three degrees of freedom, i.e. $\rho_A$,
$\rho_B$, and $r$. The different curves are related to
equations (\ref{eq:stationary}) (dot-dashed), (\ref{eq:stable-rho}) and
(\ref{e173}) (long-dashed), (\ref{e171}) and (\ref{e165})
(short-dashed), respectively.}
\label{fig6}
\end{figure}

We are now in a position to discuss the stability for different
interactions. We consider again equal particle division while
independently varying the scattering lengths $a_A = a_B < 0$ and
$a_{AB}$. The numerically computed stability diagram is shown in
figure \ref{fig6}. We shall discuss the different regions in this
contour plot in comparison with the analytic derivation.

First, we consider two attractive subsystems, $a_{AB}/b_t \leq0$, where
the coinciding centers are preferred. The computed stability condition
then follows the expression in equation (\ref{eq:stationary}), which for
$a_{AB} = 0$ reduces to the one-component stability condition.  Thus,
the attraction between $A$ and $B$ reduces the stability region. It is
amusing that $a_{A} = a_{B} = 0$ leads to the apparently identical
stability condition of $N_A a_{AB}/b_t > -0.67$. However, since $N_A =
N_B$ now the total number of pair interactings is twice as large,
i.e. $N_A^2$ compared to $N_A^2/2$.

For relatively small repulsion, $0 < a_{AB}/b_t \lesssim 0.0001$, 
the stability follows the expression in equation (\ref{eq:stable-rho}). This
corresponds to a local, but unstable minimum at $r=0$, since
equation (\ref{e173}) then also is obeyed.  These conditions are related to
structures of coinciding centers and therefore they are the same as
derived from the mean-field approximation \cite{BCP97}.

For larger repulsion, $0.0001 \lesssim a_{AB}/b_t$, 
separate center of masses of the two systems are energetically favored.  
The condition for separation is given in equation (\ref{e165}) 
which follows the computed
stability curve in the interval $0.0001 \lesssim a_{AB}/b_t < 0.0003$.
The substantial reduction of the region of mean-field stable solutions
is then defined by the difference between the curves described by
equations (\ref{e165}) and (\ref{eq:stable-rho}).

For larger repulsion $a_{AB}/b_t > 0.0003$ the stable region is larger
than found from equation (\ref{e165}), because the structure at finite $r$
is stable for relatively strong repulsion between the subsystems.
This repulsion has the effect of squeezing each of the subsystems more
than if they had been left alone as described by
equation (\ref{eq:stationary}).  Then the stable region in figure \ref{fig6}
is in fact smaller than for non-interacting subsystems, because the
barrier preventing collapse now is circumvented by the energetically
favored smaller sizes and the finite $r$.  Each subsystem therefore
finds itself with values of $\rho_{A \rm min}$ and $\rho_{B \rm min}$ on the
unstable side of the barrier preventing collapse if each were left
alone.  The mutual repulsion destabilizes the total system compared to
separate subsystems.

\section{Conclusions}
\label{sec5}

We use hyperspherical coordinates to describe a mixture of two
different components each consisting of identical bosons.
Sufficiently dilute and spatially extended systems can be rather well
approximated by use of only $s$-waves for all relative angular
momenta. From far apart only the monopole part of the interactions is
important and this is reflected in the wave function. We are then left
with a wave function only depending on the most important coordinates,
i.e. the hyperradius for each subsystem and the distance between the
two center of masses.

With a wave function depending on these three collective coordinates
and three-dimensional $\delta$-functions as the two-body interactions
we derive the corresponding Schr\"odinger equation and compute the
effective potential.  The behavior of this potential and the external
field as a function of the three collective coordinates are decisive for
the properties of the two-component system. Several features are
similar to results from the mean-field approximation.  We used the
simplest non-trivial assumptions and leave lots of room for
improvements.  Especially two-body correlations could be included in
the wave function and finite range two-body interactions with the
correct behavior could be employed.

When the interaction between particles in the two subsystems is
attractive, the potential has a minimum for coinciding centers of
mass. Then the stable structures are quite similar to those obtained
with the mean-field approximation. When the repulsion is sufficiently
strong, new stable structures arise with finite distance between the
centers.  Asymmetric particle numbers in the subsystems favor these
structures. For moderate repulsion two stable structures are
simultaneously present for centers coinciding and at a finite distance. 

New vibrational modes appear. The three normal modes essentially
correspond to relative center of mass motion with the lowest energy,
isovector and isoscalar motion with out-of-phase and in-phase
oscillations of the two subsystems as the second and third mode,
respectively.  Only the highest-lying isoscalar vibration is present
in one-component systems, where the excitation energy of $\sqrt{5}$
times the energy quantum of the harmonic trap almost precisely is
found in our computations.

New stability conditions are established. The relative center of mass
degree of freedom provides both more stable structures and reduced
stability compared to mean-field calculations.  Even when the
independent subsystems are marginally stable, a comparably strong
repulsion between the subsystems destabilizes the total system.

In conclusion, new features arise for two-component systems, i.e.
different stable structures, excitation modes, and stability
conditions.

\appendix
\section{The interaction $V_{AB}$}
\label{app:int}

We want to compute the expectation value of the interaction $V_{AB}$
in equation (\ref{e7}) over all angles, i.e. all coordinates except
$\rho_A$, $\rho_B$ and $r$. We assume that the constant angular wave
function $\Phi_0$ is normalized in the corresponding space, i.e. $\int
d\Omega |\Phi_0|^2 = 1$, where $d\Omega \equiv d\Omega_r
d\Omega_{N_A-1}\int d\Omega_{N_B-1}$ with the notation from
equation (\ref{e30}) precisely defined in \cite{sor02}.  The boson symmetry
implies that this expectation value is
\begin{eqnarray}
\label{eq:app-intAB}
  \langle \Phi_0 |V_{AB}| \Phi_0 \rangle 
 &= g_{AB}
 \int d\Omega \sum_{i=1}^{N_A}\sum_{j=N_A+1}^{N}
 \delta^{(3)}(\vec r_{i} - \vec r_{j})|\Phi_0|^2  \nonumber \\
 &=g_{AB} N_A N_B |\Phi_0|^2
 \int d\Omega \delta^{(3)}(\vec r_{N_A} - \vec r_{N_B}) \; ,
\end{eqnarray}
where the position vectors can be expressed by center of mass
coordinates and Jacobi vectors $\vec \eta$, i.e.
\begin{eqnarray}
\vec r_{N_A}=\vec R_{A}+\sqrt{\frac{N_A-1}{N_A}}\vec \eta_{N_A-1} \\
\vec r_{N_B}=\vec R_{B}+\sqrt{\frac{N_B-1}{N_B}}\vec \eta_{N_B-1} \; .
\end{eqnarray}
With $\vec r_{N_A} - \vec r_{N_B} \equiv \vec r_{\eta} - \vec r$, we
need the $\delta$-function in spherical coordinates, i.e.
\begin{eqnarray}
\delta^{(3)}(\vec r_{\eta} - \vec r)
=\frac{\delta(r_{\eta}-r)}{r_{\eta}r}\sum_\ell 
\frac{2\ell+1}{4\pi}P_\ell(\cos\theta), 
\end{eqnarray}
where $P_\ell(\cos\theta)$ are Legendre polynomials, $\theta$ is
the angle between $\vec r_\eta$ and $\vec r$, and $r_\eta=|\vec
r_\eta|$. We then get
\begin{eqnarray}
\langle \Phi_0 |V_{AB}| \Phi_0 \rangle
= g_{AB}N_A N_B|\Phi_0|^2 
\int d\Omega_{N_A-1} d\Omega_{N_B-1} 
\frac{\delta(r_{\eta}-r)}{r_{\eta}r} \; ,
\end{eqnarray}
where $r_\eta$ can be rewritten, by using $\eta_{N_A-1} = \rho_{N_A-1}
\sin \alpha_{N_A-1}$ and $\eta_{N_B-1} = \rho_{N_B-1} \sin
\alpha_{N_B-1}$, as
\begin{eqnarray}
r_{\eta}^2=&\frac{N_A-1}{N_A}\rho_A^2\sin^2\alpha_{N_A-1}
+\frac{N_B-1}{N_B}\rho_B^2\sin^2\alpha_{N_B-1}  \nonumber \\
&-2\sqrt{\frac{N_A-1}{N_A}}\sqrt{\frac{N_B-1}{N_B}}\rho_A\rho_B
\sin\alpha_{N_A-1}\sin\alpha_{N_B-1}\cos\theta_{AB}
\end{eqnarray}
where $\theta_{AB}$ is the angle between $\vec \eta_{N_A-1}$ and $\vec
\eta_{N_B-1}$.

By changing integration variable from $\theta_{AB}$ to $x \equiv
r_{\eta}-r$ we arrive at equation (\ref{eq:intAB}) where the reduced
interaction is defined by
\begin{eqnarray}
I =  \frac{\bar I }{r} \frac{4N_AN_B(\rho_A \rho_B)^{1/2}}{(N_A-1)(N_B-1)}  
 \Big(\frac{2 \Gamma(\frac{3N_A-3}{2})\Gamma(\frac{3N_B-3}{2})}
{\pi \Gamma(\frac{3N_A-6}{2})\Gamma(\frac{3N_B-6}{2})} \Big)^{1/2}
\; 
\end{eqnarray}
in terms of the basic integral
\begin{eqnarray}
\label{eq:int-I}
\fl \bar I(\rho_A,\rho_B,r) \equiv  \int d\alpha_{N_A-1}\sin\alpha_{N_A-1}
 \cos^{3N_A-7}\alpha_{N_A-1} \nonumber \\ 
\times\int d\alpha_{N_B-1}\sin\alpha_{N_B-1}\cos^{3N_B-7}\alpha_{N_B-1} \nonumber \\
\times \Theta(r_\eta^{(+)}-r)\Theta(r-r_\eta^{(-)})
\end{eqnarray}
where $\Theta$ is the Heaviside function and 
\begin{eqnarray}
r_\eta^{(\pm)}&=|\sqrt{\frac{N_A-1}{N_A}}\rho_A\sin\alpha_{N_A-1}  
\pm\sqrt{\frac{N_B-1}{N_B}}\rho_B\sin\alpha_{N_B-1}|  \\
&\approx |\rho_A\sin\alpha_{N_A-1}\pm \rho_B\sin\alpha_{N_B-1}| \; \nonumber ,
\end{eqnarray}
where we from now on shall use the last approximation valid for $N_A \gg 1$
and $N_B \gg 1$.

For further derivation we treat separately various cases of different
relative sizes $\rho_A$, $\rho_B$ and $r$.  The resulting closed
analytic expression are one-dimensional integrals of simple functions
of the angle $\alpha$, i.e.
\begin{eqnarray}
\fl
f_{\pm}(\rho_A,\rho_B,r,\alpha)   \equiv 
\frac{\sin\alpha}{3N_B-6}
\cos^{3N_A-7} \alpha [1&-&\frac{(r \pm \rho_A\sin\alpha)^2}
{\rho_B^2}]^{(3N_B-6)/2} \; .
\end{eqnarray}
The integration limits are defined by the functions
\begin{eqnarray}
 g_{\pm}(\rho_A,\rho_B,r) \equiv \arcsin[|r \pm \rho_B|/\rho_A] \; .
\end{eqnarray}
We distinguish between a number of cases.

1. [$2 \rho_B < \rho_A$ and $0 \leq r<\rho_B$] or [$\rho_B \leq \rho_A < 2
\rho_B$ and $0 \leq r<\rho_A - \rho_B$]:
\begin{eqnarray}
 \bar I(\rho_A,\rho_B,r) = \int^{g_{+}}_{0} f_{-} - \int^{g_{-}}_{0} f_{+} \; .
\end{eqnarray}

2. $2 \rho_B < \rho_A$ and $ \rho_B \leq r<\rho_A - \rho_B$:
\begin{eqnarray}
 \bar I(\rho_A,\rho_B,r) = \int^{g_{+}}_{g_{-}} f_{-}  \; .
\end{eqnarray}

3. $2 \rho_B < \rho_A$ and $\rho_A - \rho_B \leq r<\rho_A + \rho_B$:
\begin{eqnarray}
 \bar I(\rho_A,\rho_B,r) = \int^{\pi/2}_{g_{-}} f_{-}  \; .
\end{eqnarray}

4. $2 \rho_B < \rho_A$ and $\rho_A + \rho_B \leq r $:
\begin{eqnarray}
 \bar I(\rho_A,\rho_B,r) = 0  \; .
\end{eqnarray}

5. $ \rho_B \leq \rho_A < 2 \rho_B$ and $\rho_A - \rho_B \leq r< \rho_B$:
\begin{eqnarray}
 \bar I(\rho_A,\rho_B,r) = \int^{\pi/2}_{0} f_{-} - \int^{g_{-}}_{0} f_{+} \; .
\end{eqnarray}

6. $ \rho_B \leq \rho_A < 2 \rho_B$ and $\rho_B \leq r<\rho_A+ \rho_B$:
\begin{eqnarray}
 \bar I(\rho_A,\rho_B,r) = \int^{\pi/2}_{g_{-}} f_{-} \; .
\end{eqnarray}

The cases not covered are obtained by interchanging $\rho_A$ and
$\rho_B$.

The limit of $r=0$ is obtained by expansion. We first assume 
$\rho_A>\rho_B$
\begin{eqnarray}  
\frac{1}{r}\bar I &\sim
\left.\frac{\partial \bar I}{\partial r}\right|_{r=0} 
+\frac{r^2}{6}\left.\frac{\partial^3 \bar I}{\partial r^3}\right|_{r=0} \\
\left.\frac{\partial \bar I}{\partial r}\right|_{r=0}
 &=\frac{2\rho_A}{\rho_B}\int^{\arcsin(\rho_B/\rho_A)}_0 d\alpha 
\sin^2\alpha \nonumber \\
&\times  \cos^{3N_A-7}\alpha 
[1-\frac{\rho_A^2}{\rho_B^2}\sin^2\alpha]^{(3N_B-8)/2} \\
\left.\frac{\partial^3 \bar I}{\partial r^3}\right|_{r=0}
&=-6(3N_B-8)\frac{\rho_A}{\rho_B^4}
\int^{\arcsin(\rho_B/\rho_A)}_0 d\alpha
\sin^2\alpha  \nonumber \\ 
&\times \cos^{3N_A-7}\alpha
[1-\frac{\rho_A^2}{\rho_B^2}\sin^2\alpha]^{(3N_B-10)/2} \\
&+2(3N_B-8)(3N_B-10)\frac{\rho_A^3}{\rho_B^6}
\int^{\arcsin(\rho_B/\rho_A)}_0 d\alpha
\sin^4\alpha  \nonumber \\ 
&\times \cos^{3N_A-7}\alpha
[1-\frac{\rho_A^2}{\rho_B^2}\sin^2\alpha]^{(3N_B-12)/2},
\end{eqnarray}
where $\left.\frac{\partial^2 \bar I}{\partial r^2}\right|_{r=0}=0$. 
The expressions for $\rho_B>\rho_A$ are found by interchanging.

\section{Stability of $\rho$-modes}
\label{app:int2}
When we assume $m_A=m_B=m$, $N_A=N_B$, $\rho_A=\rho_B$ 
and $r=0$ the eigenvalues
for the matrix in equation (\ref{eq:matrix}) are simply $\frac{\partial^2
U_{\rm eff}}{\partial \rho_A^2} \pm \frac{\partial^2 U_{\rm eff}}{\partial
\rho_A \partial \rho_B}$. The lowest eigenvalue is negative when
\begin{eqnarray}
\label{eq:stable-eig}
  g(\rho_A) \equiv m\omega^2+\frac{27N_A^2}{4}\frac{\hbar^2}{m\rho_A^4} 
+\frac{3^{5/2}N_A^{7/2}}{\pi^{1/2}}\frac{\hbar^2}{m\rho_A^5}
(a_{A}-\frac{a_{AB}}{4}) \leq 0
\end{eqnarray}
for the solution $\rho_A=\rho_{A \rm min}$ obtained from
equation (\ref{eq:sepcon2}).  This is never possible when $a_{A} \geq a_{AB}
/4$ and the then the system is always stable.  When $a_{A} < a_{AB}$ we
continue by subtracting the function $f(\rho_{A \rm min})=0$ in
equation (\ref{eq:sepcon2}) from the inequality in equation (\ref{eq:stable-eig}). This 
immediately leads to
\begin{equation} \label{eb10}
\frac{(3N_A/2)^{1/2}}{5^{1/4}} \leq
\rho_{A \rm min} \leq -\frac{5 \cdot3^{1/2}N_A^{3/2}}{2^2\pi^{1/2}}
 \frac{a_{A}}{b_t} \equiv \rho_{Ab} \; ,
\end{equation}
where we added the inequality for the lowest possible value of
$\rho_{A \rm min}$.  Then $g(\rho_{Ab}) < 0$ is the condition for
instability, which by simple insertion results in
equation (\ref{eq:stable-rho}).  
Using instead the lower limit in equation (\ref{eb10}) we
arrive at condition (\ref{e173}) which is the stability condition for a
one-component system.


\end{document}